\documentclass[twocolumn,amsmath,amssymb]{revtex4}
\usepackage{bm}
\usepackage{graphicx}
\usepackage{epstopdf}
\usepackage{amsmath}

\begin{document}

\newcommand{\uu}[1]{\underline{#1}}
\newcommand{\pp}[1]{\phantom{#1}}
\newcommand{\be}{\begin{eqnarray}}
\newcommand{\ee}{\end{eqnarray}}
\newcommand{\ve}{\varepsilon}
\newcommand{\vs}{\varsigma}
\newcommand{\Tr}{{\,\rm Tr}}
\newcommand{\pol}{\frac{1}{2}}
\newcommand{\RR}{\rotatebox[origin=c]{180}{$\mathbb{R}$} }
\newcommand{\CC}{\rotatebox[origin=c]{180}{$\mathbb{C}$} }
\newcommand{\rr}{\mathbb{R}}
\newcommand{\Exp}{{\,\rm Exp\,}}
\newcommand{\Sin}{{\,\rm Sin\,}}
\newcommand{\Cos}{{\,\rm Cos\,}}
\newcommand{\Sinh}{{\,\rm Sinh\,}}
\newcommand{\Cosh}{{\,\rm Cosh\,}}

\title{
Time travel without paradoxes: Ring resonator as a universal paradigm for looped quantum evolutions}
\author{Marek Czachor}
\affiliation{
Katedra Fizyki Teoretycznej i Informatyki Kwantowej,
Politechnika Gda\'nska, 80-233 Gda\'nsk, Poland
}
\begin{abstract}
A ring resonator involves a scattering process where a part of the output is fed again into the input. The same formal structure is encountered in the problem of time travel in a neighborhood of a closed timelike curve (CTC). We know how to describe quantum optics of ring resonators, and the resulting description agrees with experiment. We can apply the same formal strategy to any looped quantum evolution, in particular to the time travel. The argument is in its essence a topological one and thus does not refer to any concrete geometry. 
It is  shown that the resulting paradigm automatically removes logical inconsistencies associated with chronology protection, provided all input-output relations are given by unitary maps. Examples of elementary loops and a two-loop time machine illustrate the construction. In order to apply the formalism to quantum computation one has to describe multi-qubit systems interacting via CTC-based quantum gates. This is achieved by second quantization of loops. An example of a multiparticle system, with oscillators interacting via a time machine, is explicitly calculated. However, the resulting treatment of  CTCs is not equivalent to the one proposed by Deutsch in his classic paper \cite{D}. 

\end{abstract}
\maketitle

\section{Quantum feedback loops}

Time travel is a physical problem that occurs in space-times involving closed timelike curves (CTC) \cite{Morris}. What one finds in the literature typically begins with a concrete  model of classical  space-time (by van Stockhum \cite{VS}, G\"odel \cite{G}, Taub \cite{Ta}, Newman-Unti-Tamburino \cite{NUT}, Misner \cite{M}, Gott \cite{Go,Des}, Grant \cite{Gr}...). The goal of the present paper is to shift the perspective from general relativity to quantum mechanics, and look at the time travel as a general Hilbert-space problem. Basically all the conceptual difficulties are here related to feedback loops. Loops of topological origin lead to logical vicious circles.

\begin{figure}
\includegraphics[width=8 cm]{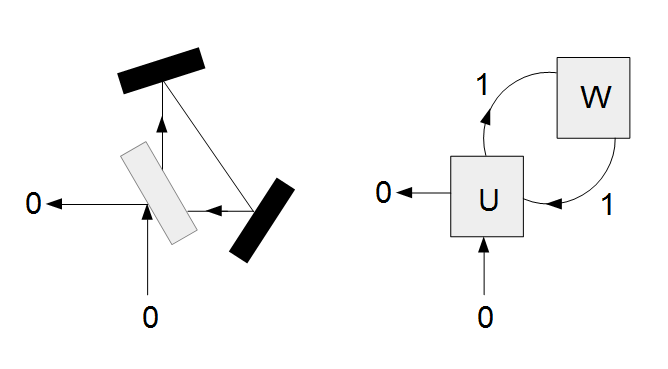}
\caption{Elementary loop. Looped interferometer (left) and its general Hilbert-space analogue. The unitary maps $U$ and $W$ act in a general Hilbert space. The two input/output `ports' of $U$ are defined by means of two arbitrary orthogonal projectors, $P_0$ and $P_1$, where the output of the subspace defined by $P_1$ is fed again into the input defined by the same projector. Restriction of $W$ to the looped subspace formally means that the map has a  block-diagonal form, $W=P_0+P_1WP_1$. Rotating the semitransparent mirror by 90 degrees we obtain a Sagnac interferometer, involving no loop. The looped interferometer is an example of a ring resonator (a ring cavity) \cite{ring}, extensively studied both experimentally and theoretically since 1960s.}
\label{Fig1}
\end{figure}

Systems whose topology leads to a feedback occur in cases much less esoteric than the time travel (see Fig.~\ref{Fig1}) but there is  no widely accepted procedure of dealing with them in quantum mechanics. Some authors suggest that the dynamics should involve nonlinear maps supplemented by consistency conditions \cite{D,R,CHB}. For example, \cite{R} reports an experimental quantum optical realization of an analogue of a Deutsch-type system \cite{D}, where a Hilbert-space nonlinearity is mimicked by means of an appropriate,  externally controlled time-dependent evolution.  On the other hand, in the context of time travel examples were given whose description reduced to a functional or path integral, and thus no Hilbert-space nonlinearity occurred \cite{B,P,Gol}. Personally sympathizing with the idea of nonlinear generalizations of quantum mechanics, I believe that looped quantum evolutions, including time machines,  can be described in a linear way.

Our guiding principle will be based on quantum optics of ring resonators (Fig.~\ref{Fig1}) \cite{ring}. Ring structures have found numerous applications in quantum information processing \cite{light,Mores,Rohde,Schreiber,He,Takeda}. 
Their theory is well grounded in experiment, so there is little doubt that quantum mechanical loops are there correctly described. 

The main idea of the proposed approach can be explained as follows. Begin with the two diagrams,
\be
\begin{array}{ccccccccc}
\psi_0(t_1) & \searrow &        &  \nearrow & \psi_0(t_2)      & \searrow &        &  \nearrow & \psi_0(t_3)                     \\
                     &                 &  U_{t_2,t_1}   &                  &        &                 &  U_{t_3,t_2}   &                  &                      \\
\psi_1(t_1) & \nearrow &        &  \searrow & \psi_1(t_2)  & \nearrow &        &  \searrow & \psi_1(t_3)
\end{array}\label{no loop'}
\ee
and
\be
\begin{array}{ccccc}
\psi_0(t_1) &  &        &   & \psi_0(t_2)\\
 & \searrow &        &  \nearrow & \\
                      &                 &  U_{t_2,t_1}   &                  &                      \\
                     & \nearrow &        &  \searrow & \\
                     & \nwarrow &         &  \swarrow & \\
                      &                 &  V     &                  &                      
\end{array}
\label{loop'}
\ee
The ``vertical" and ``horizontal axes" denote, respectively, a state-space and time. Clearly, (\ref{no loop'}) describes a composition of two time-evolution operators. Its algebraic representation reads
\be
U_{t_3,t_1}=U_{t_3,t_2}U_{t_2,t_1}.
\label{UU}
\ee

Now, what is the meaning of (\ref{loop'}), and what is its algebraic representation?  If $U_{t_2,t_1}$ is the time evolution operator from (\ref{no loop'}), then (\ref{loop'}) represents some sort of  time travel in the space of states. If the states denote position eigenstates, this is exactly the usual (quantum) time travel in a neighborhood of a CTC. This is what I call an elementary loop. Its algebraic representation is given by Theorem~1. In its essence, the construction is not geometric but topological. 

A reasoning very similar to mine, but with $U$ and $V$ explicitly constructed by path integrals,  is at the heart of  the path-integral formulation of the time machine from \cite{Gol} (technically speaking the construction from \cite{Gol} is not based on a consistency condition, as we do in the present paper, but on summation of looped propagator cycles, but the two procedures are  equivalent \cite{ver3}).  Some elements of the main idea can be also found  in \cite{Pegg,GS}, albeit in a much less general setting.
It must be stressed, though, that the `interferometer' is understood here in a very abstract sense, as any network of unitary maps, and not as some optical device. In this sense, space-time itself, if treated by path integrals, is an interferometer.

We formalize the elementary-loop consistency condition in the language of subspaces of a general Hilbert space of states. However, since our elementary loop has its optical realizations as well, we can take $U$ and $V$ from the ring-resonator literature and compare appropriate formulas,  cross-checking the general formalism. The results agree (see the Appendix), so we pass the test.

Our construction can be also seen in the context of analogue gravity and its quantum simulations.
Here, the idea is to create a medium whose acoustic or optical properties lead to field equations similar to those involving a concrete non-flat metric. Typical examples involve various aspects of acoustic  black holes (`dumb holes') \cite{Unruh,Garay,Cadoni,Lorenci,Visser98,Visser2004}, wormholes \cite{Baldovin}, brane worlds \cite{Barcelo}, FRW cosmologies \cite{Barcelo2}, cosmological constant \cite{Finazzi}, black-hole quantum teleportation \cite{Ge1}, extra dimensions \cite{Ge2}, quantum gravity \cite{Krein}, Hawking radiation in electromagnetic waveguides \cite{Unruh2},  or gravitational state vector reduction \cite{HRF}. The idea of the Universe as a `helium droplet' is discussed in great detail by Volovik \cite{Volovik}. Modern materials engineering creates another class of examples (meta-materials \cite{Sabin1603}, dc-SQUID arrays \cite{Sabin1707},  nanophotonic structures \cite{Beckenstein,Sabin1807}).  A G\"odel-type space-time can be in principle simulated by a dc-SQUID array \cite{Sabin1707,Sabin1810}.
Research in the field of analogue models of curved space-time physics has been expanding so rapidly that it is not possible to mention here all the relevant works  \cite{review}.  Still, the long list of analogue space-times does not contain ring topologies, at least not  in the sense we discuss in the paper. 

The context of analogue gravity may create an impression that what we propose is just another quantum simulation. To some extent this is true, but the goals are more ambitious. The essence of a  ring resonator is in a scattering process where a part of the output is fed again into the input. Its simplest algebraic representation corresponding to the diagram (\ref{loop'}) is $\psi_1(t_1)=V\psi_1(t_2)$. In optical realizations $V$ is an operator parametrized by the phase accumulated by light during a single cycle of the loop.  In our construction $V=P_1WP_1=W_{11}$ is  essentially arbitrary. In terms of generality,  our Theorems 1 and 2, determining an algebraic form of any looped evolution,  are comparable to the Stone theorem, stating that any one-parameter family of unitary maps satisfying $U_{\alpha+\alpha'}=U_\alpha U_{\alpha'}$ can be written as an exponent of a generator. The parameter $\alpha$ may denote time or any other parameter of a Lie group. The same level of abstractness is inherent to our $P_0$, $P_1$, $U$, $W$, $L$, and $T$. Basically, replacing the phase in $W_{11}$ by a timelike parameter we get a CTC. The results are applicable to any quantum system whose state-space topology can be represented by the right-hand sides of Figs. 1 or 2. 

Now, let us  turn to the consequences for the time travel. 

We cannot a priori exclude the possibility that a part of an input can get trapped in a looped Hilbert subspace  if one appropriately chooses the unitary maps $U$ and $W$ in Fig.~\ref{Fig1}. If this would be the case, we could invent an interferometric analogue of a black hole. To some disappointment, we will find that the resulting linear map is unitary (Theorem 1), so that anything that scatters on the system gets ultimately reflected from it. Accordingly, a looped non-dissipative system is always fully reflecting. The presence of the loop gets encoded in the structure of a scattered state. The result agrees with the fact known from optical  ring resonators. Indeed, in the ideal elementary loop topology, the intensities of input and output are equal if losses are neglected.

But the standard objection against CTCs is the grandfather paradox: Can we enter the loop, perform a time travel and kill our own grandfather? If so, how come we were born and were able to make the time travel? The solution provided  by our first theorem is simple: If you can in principle enter the loop, {\it you will not be able to do it\/}. The mouth of the wormhole will behave as an infinite potential barrier. Notice that we have obtained a general chronology protection principle \cite{Hawking,Visser}: Chronology protection is guaranteed by unitarity of quantum evolution. Details of the dynamics are irrelevant. One could not hope for a more general result.

Still, there are arguments that an evolution along the loop should not be unitary \cite{P,H95,Radu}. If this conclusion is physically correct, our version of chronology protection does not apply. 

Next, we consider the case where two loops from Fig.~\ref{Fig1} are coupled in a way shown in 
Fig.~\ref{Fig2}. The topology here is analogous to the time-machine from \cite{Gol}. Again, we find that the resulting composition of unitary maps is unitary (Theorem~2). The proofs are given in the last section. 

The two theorems are illustrated by Fock-space examples involving several interacting particles. As opposed to Deutsch, we do not invent a new quantum formalism but simply treat the unitary maps as elements of an abstract second-quantized interferometer. The case of  oscillators interacting via a time machine is explicitly computed.

Finally, we ask what happens if one destroys the loop by blocking it somehow, for example by placing there a detector. The interference at the mouth of the loop will be killed, and a putative wormhole traveler will be allowed to enter the loop. However, since the loop is in fact closed, the traveler cannot cross his own world-line (otherwise he would not be allowed to enter the loop). The phenomenon is exactly analogous to the Elitzur-Vaidman interaction-free measurement \cite{EV}, but here it becomes an ingredient of chronology protection.

We end the paper by comparison with the Deutsch approach \cite{D}. The Appendix collects some formulas from \cite{ring} and compares them with our Theorem~1. 
	
\section{How to loop a Hilbert subspace?}

Consider a general quantum dynamical problem $\psi^{\textrm{out}}=U\psi^{\textrm{in}}$ where $U$ is a unitary map (an $S$ matrix, an evolution operator $U(t,t_0)$, a quantum gate, a beam splitter, whatever). Let us split the input and the output into pairs of `ports', as represented by the diagram
\be
\begin{array}{ccccc}
\psi_0^{\textrm{in}} & \searrow &        &  \nearrow & \psi_0^{\textrm{out}}\\
                     &                 &  U   &                  &                      \\
\psi_1^{\textrm{in}} & \nearrow &        &  \searrow & \psi_1^{\textrm{out}}
\end{array}\label{no loop}
\ee
The splitting is defined by means of an arbitrary pair of orthogonal projectors, $P_0+P_1=1$, $P_0P_1=0$, 
$\psi_0^{\textrm{in}} =P_0\psi^{\textrm{in}}$, $\psi_0^{\textrm{out}}=P_0\psi^{\textrm{out}}$.  The first  goal of this paper is to give a general formula for an `elementary' loop (Fig.~\ref{Fig1}),
\be
\psi_0^{\textrm{out}}=L_{00}\psi_0^{\textrm{in}},\label{L}
\ee
obtained by looping the dynamics according to the diagram
\be
\begin{array}{ccccc}
\psi_0^{\textrm{in}} &  &        &   & \psi_0^{\textrm{out}}\\
 & \searrow &        &  \nearrow & \\
                      &                 &  U   &                  &                      \\
                     & \nearrow &        &  \searrow & \\
                     & \nwarrow &         &  \swarrow & \\
                      &                 &  W_{11}     &                  &                      
\end{array}
\label{loop}
\ee
where $W_{11}=P_1W P_1$. $W=P_0+W_{11}$  is a unitary map responsible for the evolution along the loop. The diagram implies a consistency condition for the looped subspace,
\be
\psi_1^{\textrm{in}}=W_{11}\psi_1^{\textrm{out}}.\label{ring form}
\ee
Unitarity means here that $W_{11}^*W_{11}=W_{11}W_{11}^*=P_1$. The diagram has the topology from Fig.~\ref{Fig1}. 
Notice that, in principle, the presence of the loop may change the properties the operator $U$ might have in the absence of the loop (due to a change of boundary conditions). We assume that all these possible modifications of $U$ have already been taken into account in the definition of $U$ occurring in the proof of the formula for $L_{00}$. This is not a limitation of our argument but a mathematical consistency condition. 
Now, denoting $U_{kl}=P_kUP_l$, $W_{kl}=P_kWP_l$, we obtain the following

{\it Theorem 1\/}: (Looped unitary is unlooped-unitary) Let $U$ and $W$ occurring in (\ref{loop}) be unitary, and $1-U_{11}W_{11}$ be invertible. Then, an input state is transformed into an output state by means of a linear transformation $L_{00}$ possessing the following properties:
\be
L_{00}
&=&
U_{00}
+
U_{01}W_{11}\frac{1}{1-U_{11}W_{11}}U_{10},\label{La}\\
&=&
U_{00}
+
U_{01}\frac{1}{1-W_{11}U_{11}}W_{11}U_{10},\label{Lb}\\
L_{00} &=& P_0L_{00}P_0,\\
L_{00}L_{00}^*
&=&
L_{00}^*L_{00} = P_0.
\ee

{\it Proof\/}: See Section~\ref{Proof 1}.  $\Box$

The looped composition of unitaries depicted in Fig.~\ref{Fig1} is hence itself unitary, no matter which $U$ and $W_{11}$ one takes.
Theorem~1 means that it is not possible to trap a part of the input in the loop. A looped `beam splitter' is always fully reflecting, but the fact that there exists a loop is encoded in properties of the outgoing state. In the simplest case of a $2\times 2$ matrix $U$, the operator $L_{00}$ is just a phase factor. 

\begin{figure}
\includegraphics[width=8 cm]{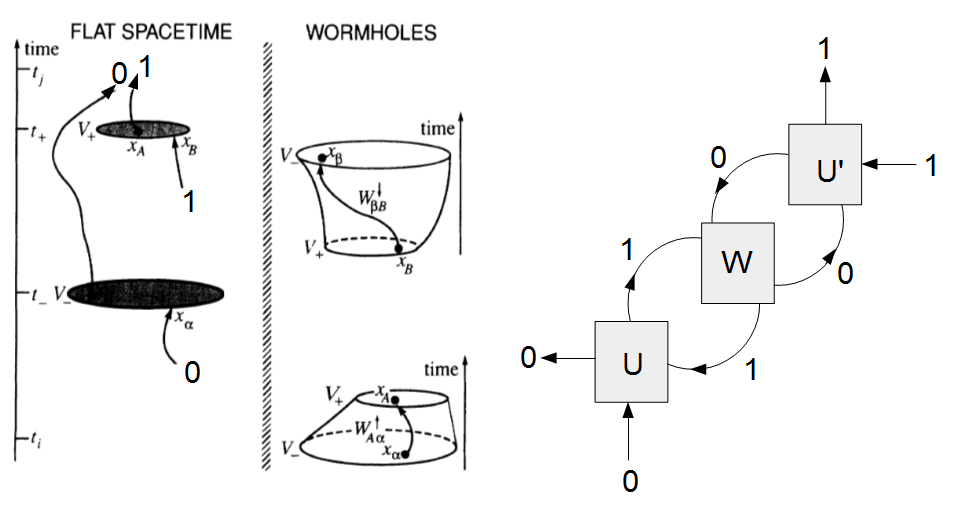}
\caption{Two coupled elementary loops. Time machine from \cite{Gol} (left) and its general Hilbert-space analogue. Here we have six input/output ports of matching dimensions, but the four ports of $W$ are looped with appropriate ports of $U$ and $U'$.}
\label{Fig2}
\end{figure}

Now consider the time-machine from the right part of Fig.~\ref{Fig2}. The diagram
\be
\begin{array}{ccccccccc}
                     & P_0\searrow &           &  \nearrow P_0  &    &                    &                &                   &             \\
                     &                 &  U      &                  &                      &                    &                &                   &             \\
                     & P_1\nearrow &           &  P_1\searrow  &                      &  \swarrow P_0  &                & \nwarrow P_0&               \\
                     &                 &           &                  &       W             &                    &                &                   &             \\
                     & P_1\nwarrow &           &  P_1\swarrow &                     &  \searrow P_0   &                &  \nearrow P_0 &               \\
                     &                 &           &                  &                      &                    &     U'        &                   &               \\
                     &                 &           &                  &                      &  P_1\nearrow    &                &  \searrow  P_1&   
\end{array}
\label{2loop0}
\ee
indicates which subspaces are looped with one another.

It might seem that the pair $P_0$, $P_1$, defining inputs and outputs of both $U$ and $U'$, could be replaced by two independent pairs, $P_0$, $P_1$, and $P_0'$, $P_1'$, respectively. However, consistency of (\ref{2loop0}) demands
\be
P_0+P_1=P_0'+P_1'=P_0+P_1'=1.
\ee
Its only solution is $P_0=P_0'$, $P_1=P_1'$.

For $W=P_0WP_0+P_1WP_1=W_{00}+W_{11}$, the system is equivalent to two separate elementary loops. If $W=P_0WP_1+P_1WP_0=W_{01}+W_{10}$ we essentially get the time machine from \cite{Gol}. Let us concentrate on the latter special case, still keeping in mind that contemporary experimental ring-resonator configurations are often much more complicated.

{\it Theorem 2\/}: Let $U$, $U'$ and $W=W_{01}+W_{10}$ occurring in diagram (\ref{2loop0}) be unitary. The diagram defines a unitary time machine $T$, $TT^* = T^*T=1$, whose explicit form reads
\be
T
&=&
U_{00}
+U_{01}W\frac{1}{1-U'_{00}WU_{11}W}U'_{00}WU_{10}
\nonumber\\
&\pp=& 
+U'_{10}W\frac{1}{1-U_{11}WU'_{00}W}U_{10}
\nonumber\\
&\pp=& 
+U_{01}W\frac{1}{1-U'_{00}WU_{11}W}U'_{01}
\nonumber\\
&\pp=& 
+
U'_{11}
+U'_{10}W\frac{1}{1-U_{11}WU'_{00}W}U_{11}WU'_{01}.\label{T}
\ee
We assume that all the operators occurring in the denominators of (\ref{T}) are invertible. 

{\it Proof\/}: See Section~\ref{Proof 2}.
$\Box$

\section{Examples}

The examples show how to apply the formalism to qubits and systems of particles. Oscillators interacting via a CTC provide one such case.  The resulting description is completely different from the one proposed by Deutsch \cite{D}.

\subsection{Looped-in-time two-dimensional rotation}
\label{SecEx1}

\begin{figure}
\includegraphics[width=8 cm]{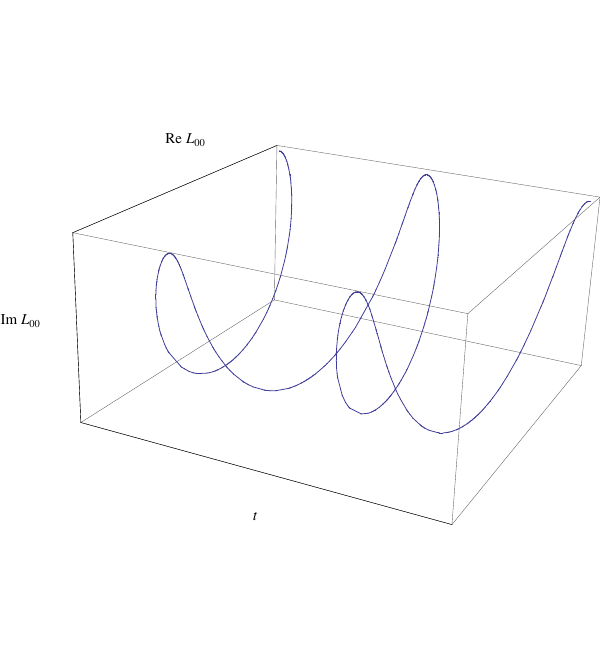}
\caption{Halfpipe phase of a looped-in-time rotation. Real and imaginary parts of the phase factor $L_{00}$ given by (\ref{L Ex1}). Here $\omega=2\pi$, $\phi=\pi/2$, $0\leq t\leq 2$.}
\label{FigEx1}
\end{figure}

Let us begin with a single qubit whose dynamics is given by a real rotation,
\be
\left(
\begin{array}{c}
\psi_{0}(t)\\
\psi_{1}(t)
\end{array}
\right)
&=&
\left(
\begin{array}{cc}
\cos\omega t & \sin\omega t\\
-\sin\omega t & \cos\omega t
\end{array}
\right)
\left(
\begin{array}{c}
\psi_{0}(0)\\
\psi_{1}(0)
\end{array}
\right).
\ee
So, here, 
\be
U&=& \left(
\begin{array}{cc}
U_{00} & U_{01}\\
U_{10} & U_{11}
\end{array}
\right)
=
\left(
\begin{array}{cc}
\cos\omega t & \sin\omega t\\
-\sin\omega t & \cos\omega t
\end{array}
\right).
\ee
Now, assume the subspace corresponding to $\psi_1$ gets looped in an elementary way (Theorem~1). The unitary operator describing the loop is a phase factor $W_{11}=e^{i\phi}$. The whole $W$ is thus given by
\be
W&=& \left(
\begin{array}{cc}
W_{00} & 0\\
0 & W_{11}
\end{array}
\right)
=
\left(
\begin{array}{cc}
1 & 0\\
0 & e^{i\phi}
\end{array}
\right).
\ee
The un-looped dynamics is effectively 1-dimensional, so $L_{00}$ is a time-dependent phase factor, 
\be
L_{00}(t)
&=&
-e^{-i\phi}
\frac{e^{i\phi}-\cos\omega t }{e^{-i\phi}-\cos\omega t }\label{L Ex1}.
\ee
Fig.~\ref{FigEx1} shows the dynamics of (\ref{L Ex1}) for  $\omega=2\pi$, $\phi=\pi/2$. 

\subsection{Two-qubit time machine}
\label{SecEx2}

The time machine is constructed from two rotations coupled by a unitary NOT operator $W$ of the form employed in Theorem~2. Let
\be
U&=& \left(
\begin{array}{cc}
U_{00} & U_{01}\\
U_{10} & U_{11}
\end{array}
\right)
=
\left(
\begin{array}{cc}
\cos\omega t & \sin\omega t\\
-\sin\omega t & \cos\omega t
\end{array}
\right),\\
U'&=& \left(
\begin{array}{cc}
U'_{00} & U'_{01}\\
U'_{10} & U'_{11}
\end{array}
\right)
=
\left(
\begin{array}{cc}
\cos\omega' t' & \sin\omega' t'\\
-\sin\omega' t' & \cos\omega' t'
\end{array}
\right),\\
W&=& \left(
\begin{array}{cc}
0 & W_{01}\\
W_{10} & 0
\end{array}
\right)
=
\left(
\begin{array}{cc}
0 & e^{i\phi}\\
e^{i\phi'} & 0
\end{array}
\right).
\ee
Inserting the above explicit forms into (\ref{T}) one finds
\be
T
=\left(
\begin{array}{cc}
\frac{\cos\omega t-e^{i(\phi'+\phi)}\cos\omega' t' }{1-e^{i(\phi'+
\phi)}\cos\omega' t' \cos\omega t }
&
\frac{e^{i\phi'}\sin\omega t \sin\omega' t' }{1-e^{i(\phi'+\phi)}\cos\omega' t' \cos\omega t }
\\
\frac{e^{i\phi}\sin\omega' t' \sin\omega t}{1-e^{i(\phi'+\phi)}\cos\omega t \cos\omega' t'}
&
\frac{\cos\omega' t'-e^{i(\phi'+\phi)}\cos\omega t}{1-e^{i(\phi'+\phi)}\cos\omega t\cos\omega' t' }
\end{array}
\right),\nonumber\\
\label{Texpl}
\ee
which is unitary, as implied by Theorem~2. To simply further analysis  assume $\omega=\omega'$, $t=t'$, $e^{i\phi}=e^{i\phi'}$, so that
\be
T(t)
&=&
\frac{(1-e^{2i\phi})\cos\omega t  }{1-e^{2i\phi}\cos^2\omega t }
+
\frac{e^{i\phi}\sin^2\omega t}{1-e^{2i\phi}\cos^2\omega t}
\sigma_1,
\ee
where $\sigma_1$ is the Pauli matrix. 
It is clear that $T(t)$ does not satisfy the theorem of Stone, and thus does not satisfy a Schr\"odinger-type equation with time-independent Hamiltonian. This happens in spite of the fact that the lopped unitaries (rotations of constant angular velocity $\omega$) did possess such generators. Formally, a time dependent Hamiltonian $H(t)=i\dot T(t)T(t)^\dag$ can nevertheless be constructed. It is unclear if it has any physical interpretation.

The solution of the eigenvalue problem $T|\pm\rangle=\tau_\pm|\pm\rangle$ is given by
\be
|\pm\rangle
&=&
\frac{1}{\sqrt{2}}
\left(
\begin{array}{c}
1\\
\pm 1
\end{array}
\right),\\
\tau_\pm
&=&
\pm e^{-i\phi}\frac{e^{i\phi}\pm   \cos\omega t}{e^{-i\phi}\pm \cos\omega t}.
\ee
It is notable that the eigenvalues have the form of elementary-loop phase factors (\ref{L Ex1}). In particular, if the two components of $|-\rangle$ denote different polarization states of a single mode of a photon, scattering of such a superposition on the time machine may be indistinguishable from interaction with the elementary loop.

\subsection{Quantum optics with the time machine}
\label{SecEx3}

The $2\times 2$ matrix $T$ from the preceding subsection can be treated as a beam splitter for quantum fields, in exact analogy to the formalism of quantum optics. Indeed, $T$ is a unitary map and as such can be written in the exponential form $T=e^{x}$ where $x^\dag=-x$.  One finds that 
\be
T
&=&
e^{i\Phi_0}e^{i\Phi_1\sigma_1},\\
e^{i\Phi_0} &=&(\tau_+\tau_-)^{1/2},\\
\Phi_1
&=&
\arctan\left(\frac{\sin\omega t}{2\sin\phi}\tan\omega t \right).
\ee
The overall phase factor $e^{i\Phi_0}$ is in this example irrelevant so let us skip it and concentrate on 
\be
T'
&=&
e^{i\arctan\left(\frac{\sin\omega t}{2\sin\phi}\tan\omega t \right)\sigma_1}.
\ee
The Jordan map $x\mapsto \hat x=\sum_{kl}a_k^\dag x_{kl}a_l$, where $[a_k,a_l^\dag]=\delta_{kl}$, is a Lie algebra isomorphism. Applying the map to the generator of $T'$, $\sigma_1\mapsto a_0^\dag a_1+a_1^\dag a_0$, one arrives at the unitary operator 
\be
\tilde T &=& e^{i\arctan\left(\frac{\sin\omega t}{2\sin\phi}\tan\omega t \right)(a_0^\dag a_1+a_1^\dag a_0)},\\
\tilde T a_k^\dag \tilde T^\dag 
&=&
\sum_l a_l^\dag T'_{lk}
\ee
acting in a two-mode Fock space. Total number of particles $a_0^\dag a_0+a_1^\dag a_1$ commutes with $\tilde T$, and 
$\tilde T|0)=\tilde T^\dag|0)=|0)$ where $a_k|0)=0$. The time machine conserves numbers of particles, as opposed to the formalism of Deutsch \cite{D} where loops map single particle states into pairs (see Section~\ref{DRev}). 

One can analogously construct fermionic extensions of looped evolutions. 

\subsection{Loops on tensor products}

The trick with the Jordan map works whenever creation and annihilation operators satisfy the Lie algebra 
$[a_k,a_l^\dag]=\delta_{kl}$. However, this algebra has various representations, differing by degrees of entanglement and other physical characteristics  \cite{PC,WC}. A two-mode representation that seems particularly useful in the context of quantum information is
\be
a_0 = a\otimes 1, \, a_1 = 1\otimes a,\, a_0^\dag = a^\dag\otimes 1, \, a_1^\dag = 1\otimes a^\dag,
\ee
$[a,a^\dag]=1$, with the vacuum $|0)=|0\rangle \otimes |0\rangle$. Physically, the Hilbert space represents two independent {\it distinguishable\/}  oscillators. The oscillators themselves are {\it not\/} bosons, a property useful for quantum coding (where order of digits cannot be ignored). Our time machine is generated by the `interaction term' 
\be
a_0^\dag a_1+a_1^\dag a_0= a^\dag\otimes a+a\otimes a^\dag,\label{30n}
\ee
whose eigenvectors
\be
\frac{1}{\sqrt{2}}(a_0^\dag\pm a_1^\dag)|0) &=& \frac{1}{\sqrt{2}}\big(|1\rangle \otimes |0\rangle\pm |0\rangle \otimes |1\rangle\big),\label{31n}
\ee
form one half of the Bell basis. The remaining two Bell states
\be
\frac{1}{\sqrt{2}}(a_0^\dag a_1^\dag\pm 1)|0) &=& \frac{1}{\sqrt{2}}\big(|1\rangle \otimes |1\rangle\pm |0\rangle \otimes |0\rangle\big),\label{32n}
\ee
or the product basis, $|0)$, $a_0^\dag |0)$, $a_1^\dag|0)$, $a_0^\dag a_1^\dag|0)$, do not possess this property.

\subsection{Elementary loop in a timelike Mach-Zehnder interferometer}

\begin{figure}
\includegraphics[width=5 cm]{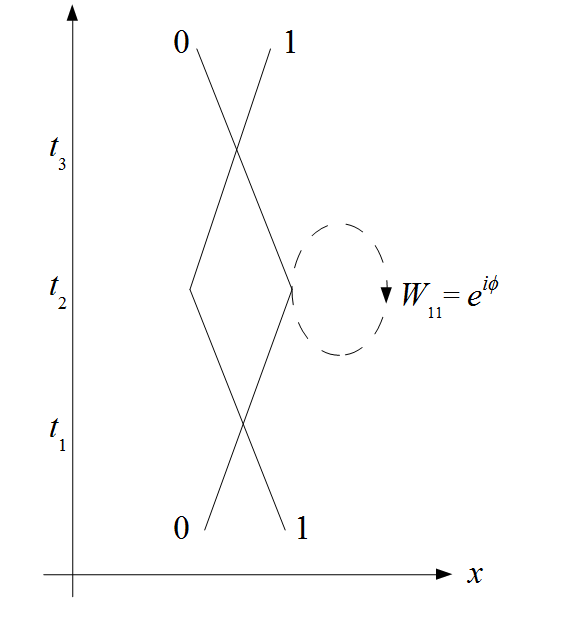}
\caption{Mach-Zehnder interferometer in time. Space-time paths of two colliding wavepackets. At half-time the wavepacket propagating to the right bounces back from the mouth of the wormhole. The scattering phase shift is determined by the structure of the CTC.}
\label{FigMZ1}
\end{figure}

The worldlines depicted in Fig.~\ref{FigMZ1} represent two scattering events. The system consists of two wavepackets scattering at $t_1$ and $t_3$. At $t_2$ the left wavepacket reflects from an infinite potential barrier whereas the right one bounces back from a mouth of a wormhole containing a CTC. Probabilities of detecting the particles at the output positions 0 or 1 are at $t_3$ given by
\be
p_k &=&|1+(-1)^kL_{00}|^2/4,\quad k=0,1.
\ee
They depend on $W_{11}$. Although the wormhole cannot be classically probed by the right particle, the structure of the CTC does influence the position-space probability at a later time $t_3$.

\section{Elitzur-Vaidman problem and chronology protection}

\begin{figure}
\includegraphics[width=8 cm]{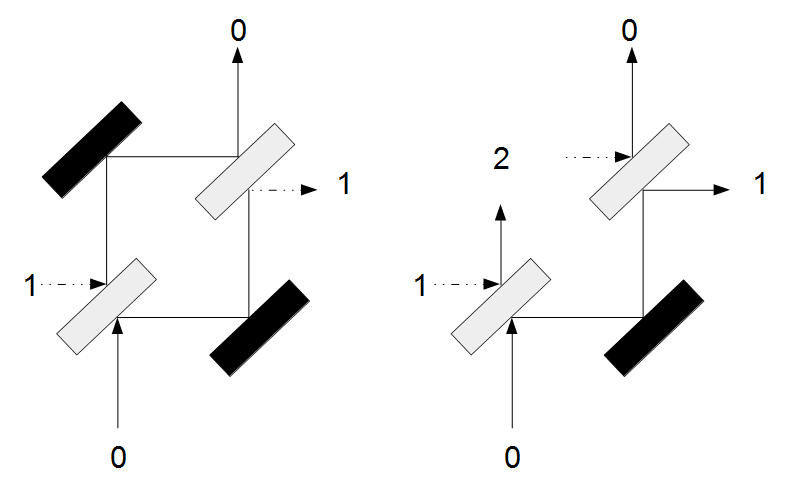}
\caption{Mach-Zehnder interferometer(left), and its opened version (right). The left system is a 2-dimensional device that acts as an identity map: 0 at the input is with certainty transmitted into 0 at the output, and 1 into 1. This is an interference effect obtained if the lengths of the two internal paths are identical. However, if we remove one mirror as shown in the right picture, the new map transfers input 0 into output 2 with probability 1/2, and into outputs 0 and 1 with probabilities 1/4. Removal of the mirror kills the interference effect at the second beam splitter so that a reflection into port 1 is no longer impossible. The same effect is found if instead of removing the mirror we place there a detector or an absorber.}
\label{Fig3}
\end{figure}

The Elitzur-Vaidman problem is related to a property of the Mach-Zehnder interferometer from the left part of Fig.~\ref{Fig3}. Namely, an amplitude representing a particle transmitted through both beam splitters destructively interferes with the one representing a particle twice reflected from them. In effect, a particle that enters through 0 has zero probability of being detected at 1. If one somehow blocks the upper internal path (by removing the mirror, or placing there an absorber or a detector) the self-interference effect is lost. A particle that enters through  0 can be detected at 1 with probability 1/4. Therefore, a detection of a particle at the output 1 means that the upper internal path was somehow tampered with. This is the essence of interaction-free measurements \cite{EV} and tests for eavesdropping in some versions of entangled-state quantum cryptography \cite{MC}.

Theorem~1 shows that the grandfather paradox is eliminated  in our formalism by the same mechanism. Indeed, consider the case of a looped wormhole. Interference at its mouth leads with certainty to reflection. The traveler cannot enter the loop and return to his world-line. However, assume that contrary to his expectation the mouth of the wormhole allowed him to start the time travel. Theorem~1 guarantees that he will not cross his world-line either. A detector or some other absorber waits for him since otherwise he would not be allowed to enter the loop. 
So, beware of quantum loops!

\section{Deutsch revisited}
\label{DRev}

The results from our two theorems can be directly compared with the formulas one obtains for ring resonators (cf.  Section 2.1.1 in \cite{ring}). The agreement is complete (see the Appendix). Experimental realizations of ring resonators include entangled states, quantum memories, heralded single photons, and so on and so forth (for a recent review see \cite{light}). All the formal ingredients needed for quantum information processing are present there. Accordingly, optical experiments seem to confirm the general discussion of loops we have given in a general Hilbert-space setting.  

But, then, what about the Deutsch formalism? The first observation is that Deutsch differently defines a looped subsystem. A system in a state $|x\rangle$ interacts via a CTC with its older version $|y\rangle$ in such a way that the entire system is described by a tensor product Hilbert space spanned by $|x\rangle\otimes |y\rangle$. So, a single-system Hilbert space is turned in a neighborhood of a CTC into a two-system Hilbert space. Deutsch's time traveler interacts with another time traveler, a copy of himself. The wormhole turns a single particle  into two particles. But why only two, and not $1+n$, corresponding to $n$ cycles of a looped evolution? Readers of Stanis{\l}aw Lem's stories remember the adventure of Ion Tichy who, trapped in a CTC, encountered a crowd of his own copies from Monday, Tuesday morning, Tuesday afternoon, Wednesday...

In my approach the time traveler remains a single object, but its state is a superposition of being inside and outside of the loop. The interaction with the wormhole is just an interference effect. My time traveler, as opposed to the Deutsch one, does not {\it interact\/} but {\it interferes\/} with the older copies of himself.  Self-interference, as opposed to self-interaction, is a linear phenomenon. The net result is the reflection from the wormhole. It must be stressed that the reflection is a consequence of taking into account all the possible cycles of the looped evolution, and not only a single cycle, as in the Deutsch proposal.

The second observation is that the notion of a looped subsystem is in the Deutsch formalism associated with a Hilbert space 
${\cal H}_2$, a part of the tensor product ${\cal H}={\cal H}_1\otimes {\cal H}_2$. Clearly, ${\cal H}_2$ is not a subspace of ${\cal H}$, so there is no projector projecting ${\cal H}$ onto ${\cal H}_2$. We have shown, however, that our formalism leads to a consistent form of multi particle looped evolutions if one preforms an appropriate Jordan-map second quantization. 
Such multi-particle systems can be used in quantum information processing involving CTCs, in exact analogy to standard quantum optical implementations of quantum gates. Various paradoxical properties of the Deutsch formalism (distinguishability of non-orthogonal states, quantum cloning, and their further implications such as faster-than-light signaling) are here absent. 

To conclude, I think that the formalism I have outlined is consistent with both quantum mechanics and experiment. The paradoxes of grandfather variety are all taken care of by quantum interference. The chronology problem is absent. Quantum mechanics of time machines remains a linear theory. It does not lead to quantum cloning or faster than light communication. Whether it has any advantages from the point of view of quantum information remains to be studied.

\section{Proofs}

\subsection{Proof of Theorem 1}
\label{Proof 1}

\subsubsection{The form of $L_{00}$}

Combining the definition of the two-port 
\be
\psi_0^{\textrm{out}}
&=&
U_{00}\psi_0^{\textrm{in}}
+
U_{01}\psi_1^{\textrm{in}},\\
\psi_1^{\textrm{out}}
&=&
U_{10}\psi_0^{\textrm{in}}
+
U_{11}\psi_1^{\textrm{in}},
\ee
with the loop,
\be
\psi_1^{\textrm{in}}
&=&
W_{11}\psi_1^{\textrm{out}},
\ee
we get, after two iterations,
\be
\psi_0^{\textrm{out}}
&=&
U_{00}\psi_0^{\textrm{in}}
+
U_{01}W_{11}\psi_1^{\textrm{out}}\label{iter 1}\\
&=&
U_{00}\psi_0^{\textrm{in}}
+
U_{01}W_{11}
\big(U_{10}\psi_0^{\textrm{in}}
+
U_{11}\psi_1^{\textrm{in}}\big)\nonumber\\
&=&
U_{00}\psi_0^{\textrm{in}}
+
U_{01}W_{11}
\big(U_{10}\psi_0^{\textrm{in}}
+
U_{11}W_{11}\psi_1^{\textrm{out}}\big)\nonumber.
\ee
Continuing the iteration one arrives at a geometric series. However, it is simpler to note the consistency condition implied by the first 
and the third equation, 
\be
\psi_1^{\textrm{out}}
&=&
U_{10}\psi_0^{\textrm{in}}
+
U_{11}W_{11}\psi_1^{\textrm{out}},
\ee
and solve it for $\psi_1^{\textrm{out}}$, assuming the inverse exists, 
\be
\psi_1^{\textrm{out}}
&=&
\frac{1}{1-U_{11}W_{11}}U_{10}\psi_0^{\textrm{in}}.
\ee
Inserting it back into (\ref{iter 1}) we obtain
\be
\psi_0^{\textrm{out}}
&=&
U_{00}\psi_0^{\textrm{in}}
+
U_{01}W_{11}\frac{1}{1-U_{11}W_{11}}U_{10}\psi_0^{\textrm{in}}\\
&=&
L_{00}\psi_0^{\textrm{in}}.
\ee
This ends the proof of the first part of the theorem.

An $n$-th term of the geometric series
\be
U_{01}W_{11}\frac{1}{1-U_{11}W_{11}}U_{10}
= \sum_{n=0}^\infty U_{01}W_{11} (U_{11}W_{11})^n U_{10}\nonumber
\ee
represents the contribution to $L_{00}$ from $n$ cycles of the looped dynamics (see \cite{ver3}).

In order to see what happens in case the series is not convergent consider the simplest case of a $2\times 2$ unitary $U$. 
 Unitarity implies $|U_{11}|\leq 1$, $|W_{11}|=1$. For $|U_{11}|< 1$ the series is convergent and $L_{00}$ is a phase factor,
\be
L_{00}
&=&
U_{00}
+
U_{01}W_{11} 
\sum_{n=0}^\infty
(U_{11}W_{11})^n
U_{10}
\nonumber\\
&=&
U_{00}
+
U_{01}W_{11}\frac{1}{1-U_{11}W_{11}}U_{10}
\nonumber\\
&=&
-W_{11}\det U\frac{1-U_{11}^* W_{11}^*}{1-U_{11}W_{11}}.
\ee
as a product of three phase factors. In the divergent case, $U_{11}W_{11}=1$, the unitarity implies $U_{10}=U_{01}=0$, and $|U_{00}|=|L_{00}|=1$. The divergence of the geometric series is therefore irrelevant since $L_{00}=U_{00}$ is a well defined phase factor. It seems that an analogous strategy will work in arbitrary dimensions, but we leave the question open.

\subsubsection{Unitarity of $L_{00}$}

Consider a unitary operator 
\be
U= \sum_{k,l=0}^1P_kUP_l=\sum_{k,l=0}^1U_{kl}.
\ee
It is convenient to represent it in a block form
\be
U&=& \left(
\begin{array}{cc}
U_{00} & U_{01}\\
U_{10} & U_{11}
\end{array}
\right)
=
\left(
\begin{array}{cc}
a & b\\
c & d
\end{array}
\right)
.
\ee
The unitarity of $U$ means 
\be
UU^*
&=&
\left(
\begin{array}{cc}
P_0 & 0\\
0 & P_1
\end{array}
\right)
=
\left(
\begin{array}{cc}
aa^*+bb^* & ac^*+bd^*\\
ca^*+db^* & cc^*+dd^*
\end{array}
\right)
\label{unit1}\\
&=&
\left(
\begin{array}{cc}
a^*a+c^*c & a^*b+c^*d\\
b^*a+d^*c & b^*b+d^*d
\end{array}
\right)=U^*U.\label{unit2}
\ee
Analogously,
\be
W&=& \left(
\begin{array}{cc}
W_{00} & 0\\
0 & W_{11}
\end{array}
\right)
=
\left(
\begin{array}{cc}
P_0 & 0\\
0 & w
\end{array}
\right),\\
ww^* &=& w^*w=P_1.\label{unit3}
\ee
Eq. (\ref{La}) can be written as 
\be
L_{00}
&=&
a+bw\frac{1}{1-dw}c,\\
L_{00}^*
&=&
a^*+c^*\frac{1}{1-w^*d^*}w^*b^*.
\ee
The rest reduces to a simple calculation, several times employing (\ref{unit1}), (\ref{unit2}), (\ref{unit3}):
\be
L_{00}L_{00}^*
&=&
aa^*
+
ac^*\frac{1}{1-w^*d^*}w^*b^*
+
bw\frac{1}{1-dw}ca^*
\nonumber\\
&\pp=&
+
bw\frac{1}{1-dw}cc^*\frac{1}{1-w^*d^*}w^*b^*
\nonumber\\
&=&
P_0-bb^*
-
bd^*\frac{1}{1-w^*d^*}w^*b^*
-
bw\frac{1}{1-dw}db^*
\nonumber\\
&\pp=&
+
bw\frac{1}{1-dw}cc^*\frac{1}{1-w^*d^*}w^*b^*
\nonumber\\
&=&
P_0-bw\frac{1}{1-dw}
\Big(
1-dd^*
-
cc^*
\Big)
\frac{1}{1-w^*d^*}w^*b^*
\nonumber\\
&=&
P_0-bw\frac{1}{1-dw}P_0\frac{1}{1-w^*d^*}w^*b^*=P_0.
\nonumber
\ee
All the explicit details of the above calculation can be found in the preprint \cite{MCv1}. 
In order to prove $L_{00}^*L_{00}=P_0$ we begin with (\ref{Lb}) and repeat similar steps. 

\subsection{Proof of Theorem 2}
\label{Proof 2}

\subsubsection{The form of $T$}

By assumption $W=W_{01}+W_{10}$ so the loop here is $\infty$-shaped (as opposed to the circle-shaped loop from Theorem~1). 
We begin with $U$, $U'$,
\be
\psi_0^{\textrm{out}}
&=&
U_{00}\psi_0^{\textrm{in}}
+
U_{01}\psi_1^{\textrm{in}}\\
\psi_1^{\textrm{out}}
&=&
U_{10}\psi_0^{\textrm{in}}
+
U_{11}\psi_1^{\textrm{in}}\\
\psi'_0{}^{\textrm{out}}
&=&
U'_{00}\psi'_0{}^{\textrm{in}}
+
U'_{01}\psi'_1{}^{\textrm{in}}\\
\psi'_1{}^{\textrm{out}}
&=&
U'_{10}\psi'_0{}^{\textrm{in}}
+
U'_{11}\psi'_1{}^{\textrm{in}},
\ee
supplemented by the loops,
\be
\psi_1^{\textrm{in}} 
&=&
W_{10}\psi'_0{}^{\textrm{out}},\\
\psi'_0{}^{\textrm{in}}
&=&
W_{01}
\psi_1^{\textrm{out}}.
\ee
The goal is to derive the transformation $T$ 
\be
\psi_0^{\textrm{out}}
&=&
T_{00}\psi_0^{\textrm{in}} + T_{01}\psi'_1{}^{\textrm{in}},\\
\psi'_1{}^{\textrm{out}}
&=&
T_{10}\psi_0^{\textrm{in}} + T_{11}\psi'_1{}^{\textrm{in}}.
\ee
What remains is similar to the proof of Theorem~1. Begin with
\be
\psi_0^{\textrm{out}}
&=&
U_{00}\psi_0^{\textrm{in}}
+
U_{01}W_{10}\psi'_0{}^{\textrm{out}}\label{38}\\
&=&
U_{00}\psi_0^{\textrm{in}}
+
U_{01}W_{10}
\big(
U'_{00}\psi'_0{}^{\textrm{in}}
+
U'_{01}\psi'_1{}^{\textrm{in}}
\big)\nonumber\\
&=&
U_{00}\psi_0^{\textrm{in}}
+
U_{01}W_{10}
\big(
U'_{00}W_{01}
\psi_1^{\textrm{out}}
+
U'_{01}\psi'_1{}^{\textrm{in}}
\big).\nonumber
\ee
The consistency condition reads
\be
\psi'_0{}^{\textrm{out}}
&=&
U'_{00}W_{01}
\psi_1^{\textrm{out}}
+
U'_{01}\psi'_1{}^{\textrm{in}}.\label{39}
\ee
Inserting
\be
\psi_1^{\textrm{out}}
&=&
U_{10}\psi_0^{\textrm{in}}
+
U_{11}W_{10}\psi'_0{}^{\textrm{out}}
\ee
into (\ref{39}) we obtain
\be
\psi'_0{}^{\textrm{out}}
=
U'_{00}W_{01}
\big(
U_{10}\psi_0^{\textrm{in}}
+
U_{11}W_{10}\psi'_0{}^{\textrm{out}}
\big)
+
U'_{01}\psi'_1{}^{\textrm{in}},\nonumber\\
\ee
which can be solved for $\psi'_0{}^{\textrm{out}}$. Indeed,
\be
\big(1-U'_{00}W_{01}U_{11}W_{10}\big)
\psi'_0{}^{\textrm{out}}
=
U'_{00}W_{01}
U_{10}\psi_0^{\textrm{in}}
+
U'_{01}\psi'_1{}^{\textrm{in}},\nonumber
\ee
so assuming $1-U'_{00}W_{01}U_{11}W_{10}$ is invertible, and returning to  
(\ref{38}), we find
\be
\psi_0^{\textrm{out}}
&=&
U_{00}\psi_0^{\textrm{in}}
\nonumber\\
&\pp=&+
U_{01}W_{10}
\frac{1}{1-U'_{00}W_{01}U_{11}W_{10}}
U'_{00}W_{01}
U_{10}\psi_0^{\textrm{in}}
\nonumber\\
&\pp=&+
U_{01}W_{10}
\frac{1}{1-U'_{00}W_{01}U_{11}W_{10}}
U'_{01}\psi'_1{}^{\textrm{in}}\\
&=&
T_{00}\psi_0^{\textrm{in}} + T_{01}\psi'_1{}^{\textrm{in}}.
\ee
We can skip the indices in $W$, arriving at the first and the third terms of (\ref{T}),
\be
T_{00}
&=&
U_{00}+U_{01}W
\frac{1}{1-U'_{00}WU_{11}W}
U'_{00}W
U_{10},\\
T_{01}
&=&
U_{01}W
\frac{1}{1-U'_{00}WU_{11}W}
U'_{01}.
\ee
Now, consider
\be
\psi'_1{}^{\textrm{out}}
&=&
U'_{10}W_{01}
\psi_1^{\textrm{out}}
+
U'_{11}\psi'_1{}^{\textrm{in}}\label{46}\\
&=&
U'_{10}W_{01}
\big(
U_{10}\psi_0^{\textrm{in}}
+
U_{11}\psi_1^{\textrm{in}}
\big)
+
U'_{11}\psi'_1{}^{\textrm{in}}\nonumber\\
&=&
U'_{10}W_{01}
\big(
U_{10}\psi_0^{\textrm{in}}
+
U_{11}W_{10}\psi'_0{}^{\textrm{out}}
\big)
+
U'_{11}\psi'_1{}^{\textrm{in}}.\nonumber
\ee
The consistency condition is
\be
\psi_1^{\textrm{out}}
&=&
U_{10}\psi_0^{\textrm{in}}
+
U_{11}W_{10}\psi'_0{}^{\textrm{out}}.\label{47}
\ee
Inserting 
\be
\psi'_0{}^{\textrm{out}}
&=&
U'_{00}W_{01}
\psi_1^{\textrm{out}}
+
U'_{01}\psi'_1{}^{\textrm{in}}
\ee
into (\ref{47}), we compute $\psi_1^{\textrm{out}}$ from
\be
\big(
1-U_{11}W_{10}U'_{00}W_{01}
\big)
\psi_1^{\textrm{out}}
=
U_{10}\psi_0^{\textrm{in}}
+
U_{11}W_{10}
U'_{01}\psi'_1{}^{\textrm{in}},
\nonumber
\ee
and insert the result into (\ref{46}). Finally,
\be
\psi'_1{}^{\textrm{out}}
&=&
U'_{10}W
\frac{1}{1-U_{11}WU'_{00}W}U_{10}\psi_0^{\textrm{in}}
\nonumber\\
&\pp=&
+
U'_{10}W\frac{1}{1-U_{11}WU'_{00}W}
U_{11}W
U'_{01}\psi'_1{}^{\textrm{in}}
\nonumber\\
&\pp=&
+
U'_{11}\psi'_1{}^{\textrm{in}}\nonumber\\
&=&
T_{10}\psi_0^{\textrm{in}} + T_{11}\psi'_1{}^{\textrm{in}}
\ee
reconstruct the second and the fourth lines of (\ref{T}). The whole formula (\ref{T}) follows from
\be
T=\sum_{k,l=0}^1P_kTP_l=\sum_{k,l=0}^1T_{kl}.
\ee

\subsubsection{Unitarity of $T$}

Let $X=UW$, $X'=U'W$. The maps are unitary. The blocks are related by
\be
X_{00}
&=&
P_0UWP_0=P_0UP_1W=U_{01}W,\\
X_{01}
&=&
P_0UWP_1=P_0UP_0W=U_{00}W,\\
X_{10}
&=&
P_1UWP_0=P_1UP_1W=U_{11}W,\\
X_{11}
&=&
P_1UWP_1=P_1UP_0W=U_{10}W,
\ee
and analogously for $X'$. Rewriting $T$ by means of $X$ and $X'$, and defining $S=TW$ we ultimately obtain a form which is more convenient for the proof (unitarity of $S$ implies the one of $T$), 
\be
S
&=&
X_{01}
+X_{00}\frac{1}{1-X'_{01}X_{10}}X'_{01}X_{11}
\nonumber\\
&\pp=& 
+X'_{11}\frac{1}{1-X_{10}X'_{01}}X_{11}
+X_{00}\frac{1}{1-X'_{01}X_{10}}X'_{00}
\nonumber\\
&\pp=& 
+
X'_{10}
+X'_{11}\frac{1}{1-X_{10}X'_{01}}X_{10}X'_{00}.
\ee
Denote,
\be
X
&=&
\left(
\begin{array}{cc}
X_{00} & X_{01} \\
X_{10} & X_{11}
\end{array}
\right)
=
\left(
\begin{array}{cc}
a & b \\
c & d
\end{array}
\right),
\\
X'
&=&
\left(
\begin{array}{cc}
X'_{00} & X'_{01} \\
X'_{10} & X'_{11}
\end{array}
\right)
=
\left(
\begin{array}{cc}
a' & b' \\
c' & d'
\end{array}
\right).
\ee
In this notation
\be
S
&=&
\left(
\begin{array}{cc}
S_{00} & S_{01} \\
S_{10} & S_{11}
\end{array}
\right)
=
\left(
\begin{array}{cc}
A & B \\
C  & D
\end{array}
\right)
\\
&=&
\left(
\begin{array}{cc}
a\frac{1}{1-b'c}a' & b+a\frac{1}{1-b'c}b'd
 \\
c'+d'\frac{1}{1-cb'}ca'  & d'\frac{1}{1-cb'}d
\end{array}
\right).\label{S}
\ee
We have to prove that $S$ is unitary whenever $X$ and $X'$ are unitary. 
But first, let us have a look at
\be
S^*
&=&
\left(
\begin{array}{cc}
A^* & C^* \\
B^*  & D^*
\end{array}
\right)
\nonumber\\
&=&
\left(
\begin{array}{cc}
a'^*\frac{1}{1-c^*b'^*}a^* & c'^*+a'^*c^*\frac{1}{1-b'^*c^*} d'^*\\
b^*+d^*b'^*\frac{1}{1-c^*b'^*}a^*  & d^*\frac{1}{1-b'^*c^*}d'^*
\end{array}
\right)
\nonumber\\
&=&
\left(
\begin{array}{cc}
a'^*\frac{1}{1-c^*b'^*}a^* & c'^*+a'^*\frac{1}{1-c^*b'^*} c^*d'^*\\
b^*+d^*\frac{1}{1-b'^*c^*}b'^*a^*  & d^*\frac{1}{1-b'^*c^*}d'^*
\end{array}
\right).\nonumber\\
\label{S*}
\ee
Comparing (\ref{S*}) with (\ref{S})
we observe that $S^*$ has the same form as $S$ if one interchanges $X$ and $X'^*$. Since $X$ and $X'$ are arbitrary unitary operators, if we manage to prove  $SS^*=1$ then $S^*S=1$ will be obtained just by $X\leftrightarrow X'^*$.
The proof of unitarity of $S$ (and thus of $T$) reduces to checking that
\be
AA^*+BB^* &=& P_0,\label{AA}\\
AC^*+BD^* &=& 0,\label{AC}\\
CC^*+DD^* &=& P_1.\label{CC}
\ee
All the three proofs are similar to the one we have given for the case of $L$ from Theorem~1. 
So, let us outline the one for (\ref{CC}), leaving the remaining ones as exercises for the readers.
The conditions to be used are (\ref{unit1}), (\ref{unit2}), together with their primed versions.
Then
\be
(\ref{CC})
&=&
c'c'^*
\nonumber\\
&\pp=&
+c'a'^*c^*\frac{1}{1-b'^*c^*} d'^*
\nonumber\\
&\pp=&
+
d'\frac{1}{1-cb'}ca'  c'^*
\nonumber\\
&\pp=&
+
d'\frac{1}{1-cb'}ca'  a'^*c^*\frac{1}{1-b'^*c^*} d'^*
\nonumber\\
&\pp=&
+
d'\frac{1}{1-cb'}d
d^*\frac{1}{1-b'^*c^*}d'^*
\nonumber\\
&=&
P_1-d'\frac{1}{1-cb'}(1-cb')(1-b'^*c^*)\frac{1}{1-b'^*c^*}d'^*
\nonumber\\
&\pp=&
-d'\frac{1}{1-cb'}(1-cb')b'^*c^*\frac{1}{1-b'^*c^*} d'^*
\nonumber\\
&\pp=&
-
d'\frac{1}{1-cb'}cb'  (1-b'^*c^*)\frac{1}{1-b'^*c^*}d'^*
\nonumber\\
&\pp=&
+
d'\frac{1}{1-cb'}c(P_0-b'  b'^*)c^*\frac{1}{1-b'^*c^*} d'^*
\nonumber\\
&\pp=&
+
d'\frac{1}{1-cb'}d
d^*\frac{1}{1-b'^*c^*}d'^*
\nonumber\\
&=&
P_1
-d'\frac{1}{1-cb'}P_0\frac{1}{1-b'^*c^*}d'^*
=P_1,
\ee
which we had to demonstrate.

\section*{Appendix: Elementary loop from \cite{ring}}

Let us compare the analysis of a single ring resonator given in \cite{ring} with our general formulas for elementary-loop map $L$. 
In the notation from \cite{ring} the beam splitter is given by (Eq. 2.1 in \cite{ring})
\be
\left(
\begin{array}{c}
E_{t1}\\
E_{t2}
\end{array}
\right)
&=&
\left(
\begin{array}{cc}
t & \kappa\\
-\bar\kappa & \bar t
\end{array}
\right)
\left(
\begin{array}{c}
E_{i1}\\
E_{i2}
\end{array}
\right),
\ee
with $|t|^2+|\kappa|^2=1$. 
This corresponds to our
\be
\left(
\begin{array}{c}
\psi_0^{\textrm{out}}\\
\psi_1^{\textrm{out}}
\end{array}
\right)
&=& \left(
\begin{array}{cc}
U_{00} & U_{01}\\
U_{10} & U_{11}
\end{array}
\right)
\left(
\begin{array}{c}
\psi_0^{\textrm{in}}\\
\psi_1^{\textrm{in}}
\end{array}
\right),
\ee
with $U^*U=UU^*=1$.
The resonator consists of a ring of radius $r$. Our feedback formula 
\be
\psi_1^{\textrm{in}}=W_{11}\psi_1^{\textrm{out}},\label{ring form'}
\ee
is represented in \cite{ring} by 
\be
E_{i2}=\alpha e^{i\theta}E_{t2},\quad \theta=\omega L/c,
\ee
where $L=2\pi r$. $\omega$, $c$ are, respectively, a mode frequency and a velocity of light in the resonator. $\alpha$ is a damping coefficient. Assuming no losses, as we do in the present paper, we should put $\alpha=1$. The solution, given by Eqs. (2.6)-(2.8) in \cite{ring}, assuming $E_{i1}=1$, reads
\be
E_{t1}
&=&
\frac{-\alpha+te^{-i\theta}}{-\alpha \bar t+e^{-i\theta}},\\
E_{i2}
&=&
\frac{-\alpha\bar\kappa}{-\alpha \bar t+e^{-i\theta}},\\
E_{t2}
&=&
\frac{-\bar\kappa}{1-\alpha \bar t e^{i\theta}}.
\ee
In our case
\be
\psi_1^{\textrm{out}}
&=&
L_{10}\psi_0^{\textrm{in}}\\
&=&
\frac{1}{1-U_{11}W_{11}}U_{10}\psi_0^{\textrm{in}}
\\
&=&
\frac{1}{1-\bar t\alpha e^{i\theta}}(-\bar\kappa )\psi_0^{\textrm{in}}=E_{t2},
\ee
since $\psi_0^{\textrm{in}}=E_{i1}=1$. 
\be
\psi_0^{\textrm{out}}
&=&
\left(
U_{00}
+
U_{01}W_{11}\frac{1}{1-U_{11}W_{11}}U_{10}
\right)
\psi_0^{\textrm{in}}\\
&=&
t
+
\kappa \alpha e^{i\theta}\frac{1}{1-\bar t\alpha e^{i\theta}}(-\bar\kappa )
\\
&=&
\frac{t-(|t|^2+|\kappa|^2)\alpha e^{i\theta}}{1-\bar t\alpha e^{i\theta}}=E_{t1},
\ee
in full agreement with \cite{ring}.


\begin{thebibliography}{99}

\bibitem{D}D. Deutsch, Phys. Rev. D {\bf 44}, 3197 (1991).

\bibitem{Morris}M. S. Morris, K. S. Thorne, and U. Yurtsever,  Phys. Rev. Lett. {\bf 61}, 1446 (1988).

\bibitem{VS}W. J. Van Stockum, Proc. Roy. Soc. Edin. {\bf 57}, 135 (1937).
\bibitem{G}K. G\"odel, Rev. Mod. Phys. {\bf 21}, 447 (1949).
\bibitem{Ta}A. H. Taub, Ann. Math. {\bf 53}, 472 (1951).
\bibitem{NUT}E. T. Newman, L. Tamburino, and T. J. Unti, J. Math. Phys. {\bf 4}, 915 (1963).
\bibitem{M}C. W. Misner, in {\it Relativity Theory and Astrophysics I. Relativity and Cosmology\/}, ed. J. Ehlers, pp. 160-169, Providence (1967).
\bibitem{Go}J. R. Gott III, Phys. Rev. Lett. {\bf 66}, 1126 (1991).
\bibitem{Des}S. Deser, R. Jackiw, and G. 't Hooft, Phys. Rev. Lett. {\bf 68}, 267 (1992).
\bibitem{Gr}J. D. E. Grant, Phys. Rev. D {\bf 47}, 2388 (1993).

\bibitem{R} M. Ringbauer et al., Nat. Comm. {\bf 5}, 4145 (2014) .
\bibitem{CHB}C. H. Bennett et al., Phys. Rev. Lett. {\bf 103}, 170502 (2009).

\bibitem{B}D. G. Boulware, Phys. Rev. D. {\bf 46}, 4421 (1992).
\bibitem{P}H. D. Politzer, Phys. Rev. D. {\bf 46}, 4410 (1992);  Phys. Rev. D. {\bf 49}, 3981 (1994).
\bibitem{Gol}D. S. Goldwirth et al., Phys. Rev. D. {\bf 49}, 3951 (1994).


\bibitem{ring}D. G. Rabus, {\it Integrated Ring Resonators. The Compendium\/}, Springer, Berlin (2007).

\bibitem{light}L. Caspani et al., Light: Science \& Applications {\bf 6}, e17100 (2017); doi: 10.1038/lsa.2017.100
\bibitem{Mores}K. R. Motes, A. Gilchrist, J. P. Dowling, and P. P. Rohde,  Phys. Rev. Lett. {\bf 113}, 120501 (2014).
\bibitem{Rohde}P. P. Rohde,  Phys. Rev. A {\bf 91}, 012306 (2015).
\bibitem{Schreiber}A. Schreiber et al., Science {\bf 336}, 55 (2012).
\bibitem{He}Y. He et al., Phys. Rev. Lett. {\bf 118}, 190501 (2017).
\bibitem{Takeda}S. Takeda and A. Furusawa,  Phys. Rev. Lett. {\bf 119}, 120504 (2017).
\bibitem{ver3}In an earlier version of the present paper, M. Czachor,  arXiv:1805.12129v4 [quant-ph], proofs employing a summation of cycles of looped evolutions were used. As expected, the results were identical to our Theorems 1 and 2. The present proofs are more straightforward, and do not require additional `physical' assumptions about the physical process itself. It is nevertheless very instructive to prove the theorems in both ways.

\bibitem{Pegg} T. Pegg, in {\it Times' arrows, quantum measurement and superluminal behavior\/}, eds. D. Mugnai, A. Ranfagni and L. S. Schulman, p. 113, CNDR, Roma (2001); arXiv:quant-ph/0506141.
\bibitem{GS}D. M. Greenberger and K. Svozil, in {\it Between Chance and Choice\/}, eds.  H. Atmanspacher and R. Bishop, p. 293, Imprint Academic, Thorverton (2002); arXiv:quant-ph/0506027 (2005).

\bibitem{H95}S. W. Hawking, Phys. Rev. D {\bf 52}, 5681 (1995).
\bibitem{Radu}E. Radu, Phys. Lett. A {\bf 247}, 207 (1998).


\bibitem{Unruh}W. G. Unruh, Phys. Rev. Lett. {\bf 46} 1351 (1981).
\bibitem{Garay} L.J. Garay, J.R. Anglin, J.I. Cirac  and P. Zoller,  Phys. Rev. Lett. {\bf 85}, 4643 (2000).
\bibitem{Cadoni}M. Cadoni and S. Mignemi,  Phys. Rev. D {\bf 72}, 084012, (2005).
\bibitem{Lorenci}V.A. De Lorenci, R.   Klippert, R. and  Y.N. Obukhov, Phys. Rev. D {\bf 68}, 061502 (2003).
\bibitem{Visser98}M. Visser, Class. Quantum Grav. {\bf 15}, 1767 (1998).
\bibitem{Visser2004}C. Barcel{\'o}, S. Liberati, S. Sonego and  M. Visser, New J. Phys. {\bf 6}, 186 (2004 ).
\bibitem{Baldovin}F. Baldovin, M.  Novello, S.E. Perez Bergliaffa, and J.M. Salim,  Class. Quantum Grav. {\bf 17}, 3265 (2000).
\bibitem{Barcelo}C. Barcel{\'o}  and A. Campos,  Phys. Lett. B {\bf 563}, 217 (2003).
\bibitem{Barcelo2}C. Barcel{\'o}, S. Liberati and M. Visser,  Int. J. Mod. Phys. D {\bf 12}, 1641 (2003).
\bibitem{Finazzi}S. Finazzi, S. Liberati and L. Sindoni,  Phys. Rev. Lett. {\bf 108}, 071101 (2012)
\bibitem{Ge1}X.-H. Ge and  Y.-G. Shen, Phys. Lett. B {\bf 623}, 141 (2005).
\bibitem{Ge2}X.-H. Ge and  S.-W. Kim, Phys. Lett. B {\bf 652}, 349 (2007).
\bibitem{Krein}G. Krein, G. Menezes and N.F. Svaiter, Phys. Rev. Lett. {\bf 105}, 131301 (2010).
\bibitem{Unruh2}R. Sch{\"u}tzhold and W.G. Unruh, Phys. Rev. Lett. {\bf 95}, 031301 (2005).
\bibitem{HRF}R. Howl, R. Penrose, and I. Fuentes, New J. Phys. {\bf 21}, 043047 (2019).
\bibitem{Volovik}G. E. Volovik, {\it The Universe in a Helium Droplet\/}, Clarendon Press, Oxford (2003).


\bibitem{Sabin1603}C. Sab{\'\i}n, Phys. Rev. D {\bf 94}, 081501 (2016).
\bibitem{Sabin1807}C. Sab{\'\i}n, Universe {\bf 4}, 115 (2018).
\bibitem{Beckenstein}R. Bekenstein et al.,  Nature Phot. {\bf 11}, 664 (2017).

\bibitem{Sabin1707}C. Sab{\'\i}n, New J. Phys. {\bf 20}, 053028 (2018).
\bibitem{Sabin1810}G. Mart\'\i n-V\'azquez and  C. Sab{\'\i}n, arXiv:1810.05124 [quant-ph] (2018).

\bibitem{review}C. Barcel{\'o}, S. Liberati and  M. Visser, Living Rev. Rel. {\bf 8}12 (2005); the updated preprint 
arXiv:gr-qc/0505065  (version from 2011) includes 702 references.



\bibitem{Hawking}S. W. Hawking, Phys. Rev. D {\bf 46}, 603 (1992).
\bibitem{Visser}M. Visser, Phys. Rev. D {\bf 47}, 554 (1993).
\bibitem{PC}M. Paw{\l}owski and M. Czachor,  Phys. Rev. A {\bf 73}, 042111 (2006).
\bibitem{WC}M. Wilczewski and M. Czachor,  Phys. Rev. A {\bf 80}, 013802 (2009).
\bibitem{EV}A. C. Elitzur and L. Vaidman, Found. Phys. {\bf 23}, 987 (1993).
\bibitem{MC}M. Czachor, Phys. Lett. A {\bf 257}, 107 (1999); arXiv:quant-ph/9812030.
\bibitem{MCv1}M. Czachor, arXiv:1805.12129v1 [quant-ph]

\end{thebibliography}
\end{document}